\input{aipcheck}

\documentclass[
    ,final            
  ,numberedheadings 
  ]
  {aipproc}

\layoutstyle{8x11single}

\begin{document}

\title{Stellar Encounters with Multiple Star Systems and the Blue Straggler Problem}


\author{Nathan Leigh}{
  address={Department of Physics and Astronomy, McMaster University, 1280 Main St. W., Hamilton, ON, L8S 4M1, Canada}
}
\author{Alison Sills}{
  address={Department of Physics and Astronomy, McMaster University, 1280 Main St. W., Hamilton, ON, L8S 4M1, Canada}
}

\begin{abstract}
We present a technique to identify the most probable dynamical
formation scenario for an observed binary or triple system containing
one or more merger products or, alternatively, to rule out the
possibility of a dynamical origin.  Our method relies on an analytic
prescription for energy conservation during stellar encounters.
With this, observations of the multiple star system containing the
merger
product(s) can be used to work backwards in order
to constrain the initial orbital energies of any single, binary or
triple
systems that went into the encounter.  The initial semi-major axes of
the orbits provide an
estimate for the collisional cross section and therefore the
time-scale for the encounter to occur in its host cluster.

We have applied our analytic prescription to observed binary and
triple systems containing blue stragglers, in particular the triple
system S1082 in M67 and the period distribution of the blue
straggler binaries in NGC 188.  We have shown that both S1082 and
most of the blue straggler binaries in NGC 188 could have a
dynamical origin, and that encounters involving triples are a
significant contributor to blue straggler populations in old open clusters.
In general, our results suggest that encounters involving triples
could make up a
significant fraction of those dynamical interactions that result in
stellar
mergers, in particular encounters that produce multiple star systems
containing one or more blue stragglers.
\end{abstract}

\maketitle

\section{Introduction}

It has been known for some time that encounters, and even direct
collisions, can occur frequently between stars in dense stellar
systems
\citep[e.g.][]{hills76, hut83, leonard89}.  In the cores of globular
clusters (GCs), the time between collisions
involving two single stars can
be much shorter than the cluster lifetime \citep{leonard89}.
The time between encounters involving binary stars can be
considerably shorter still given their much larger cross sections for
collision.  In globular and, especially, open clusters (OCs) with high
binary fractions, mergers are thought to occur frequently during
resonant interactions involving binaries \citep[e.g.][]{leonard92}.
What's more, collision products have a significant probability of
undergoing more than one
collision during a given single-binary or binary-binary interaction
since the initial impact is expected to result in shock heating
followed by adiabatic expansion, increasing the cross section for a
second collision to occur \citep[e.g.][]{fregeau04}.

Several types of stars whose origins remain a mystery are speculated
to be the products of stellar mergers.  Blue stragglers (BSs) in
particular are thought to be produced via the addition of fresh
hydrogen to low-mass main-sequence (MS) stars
\cite[e.g.][]{sills01}.  Recent 
evidence has shown that, whatever the dominant BS formation
mechanism(s) operating in both globular and open clusters, it
is likely to in some way depend on binary stars \citep{knigge09,
  mathieu09}.  The currently favored mechanisms include collisions
during single-binary and binary-binary encounters
\citep[e.g.][]{leonard89}, mass transfer
between the components of a binary system \citep[e.g.][]{chen08a,
  chen08b} and the coalescence of two
stars in a close binary due to perturbations from an orbiting
triple companion \citep[e.g.][]{eggleton06, perets09}.  Regardless of
the dominant BS formation mechanism(s) operating in dense star
clusters,
dynamical interactions should play at least some role.  For example,
even if blue stragglers are formed as a result of binary evolution
processes such as mass transfer, the progenitor binaries themselves
should have been affected by at least one dynamical
interaction over the course of their lifetime.

In these proceedings, we introduce an analytic technique to
constrain the most probable dynamical origin of an observed binary or
triple system containing one or more merger products.  Provided the
observed system is found within a moderately dense cluster
environment with binary and/or triple fractions of at least a few
percent, the probability is often high that it formed from a merger
during
an encounter involving one or more binary or triple stars.  In
Section~\ref{method}, we describe a prescription for energy conservation
during individual stellar encounters and outline the process for
applying our technique.  Specifically, we present a step-by-step
methodology to evaluate whether or not an assumed dynamical history
could have realistically produced an observed system and describe how
to determine the most probable dynamical formation scenario.
In Section~\ref{results}, we apply
our technique to a few observed binary and triple systems thought to
contain merger
products, in particular a triple system that is thought to contain two
BSs and
the peculiar period distribution of the BS binary
population in NGC 188.  We discuss the implications of our results in
Section~\ref{discussion}.

\section{Method} \label{method}

In this section, we describe a technique to constrain the dynamical
origin of an observed binary or triple containing one or more BSs.
Alternatively, our technique can be used to altogether rule out a
dynamical origin.  We will limit the
discussion to typical interactions thought to occur in old open clusters.

\subsection{Finding the Dominant Encounter Type} \label{encounters}

\begin{figure}
  \includegraphics[height=.3\textheight]{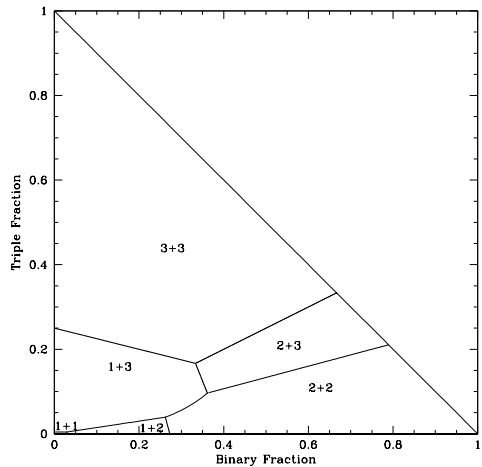}
\caption{Parameter space in the binary
  fraction-triple fraction plane for
which each of the various encounter types dominate.  Boundaries
between regions are indicated by solid lines, each segment of which is
obtained by equating two particular encounter rates using
Equation 12 of \cite{leonard89} and the relevant cross sections derived
using Equation 6.}  
\label{fig:fb-ft}
\end{figure}

We begin by assuming that an observed system was formed
directly from a dynamical interaction (or sequence of
interactions).  The first step in our procedure is to determine which
type of dynamical interaction(s) is the most
likely to have produced the observed binary or triple.  First, the
analytic rates for single-single (1+1), single-binary (1+2),
single-triple (1+3), binary-binary (2+2), binary-triple (2+3) and
triple-triple (3+3) encounters can be
compared to obtain a rough guide as to which of these encounter
types will dominate in a given cluster.  The total rate of
encounters in a cluster core is given by Equation 12 of
\cite{leonard89}.  From this, expressions can be found for each of
the different types of encounters (see \cite{leigh10} for more
details).  Assuming for simplicity that
the relative velocity at infinity is roughly equal for all types of
encounters, the rates for two types of encounters can be compared to
find the parameter space 
for which one type of encounter will dominate over another.  We let
$f_b$ and $f_t$ represent the fraction of objects that are binaries
and triples, respectively.  These relations can then be plotted in
the $f_b-f_t$ plane in order to partition the parameter space for
which each of the various encounter types will occur with the greatest
frequency.  This is shown in Figure~\ref{fig:fb-ft}.  Given a
cluster's binary and triple fractions, this provides a
simple means of finding the type of encounter that will occur with the
greatest frequency.  

\subsection{Conservation of Energy} \label{energy}

We have assumed that the observed binary or triple containing the
merger product(s) has a
dynamical origin.  Therefore, the observed parameters of the system
provide the final distribution of energies left-over
after the interaction(s).  Once we have found the dominant encounter type
for the cluster being considered, this will serve as a guide in
choosing an appropriate encounter scenario for the observed
binary or triple.  With a specific dynamical scenario in mind, we can
work backwards using energy conservation and observations of the host
cluster to constrain the initial
energies going into the encounter(s).  This provides an estimate for the
initial orbital energies and therefore the semi-major axes of any binaries
or triples going into the interaction.  This in turn gives the cross
section for collision and hence the time required for the hypothesized
interaction(s) to occur. 

Since the formation event must have happened in the last
$\tau_{BS}$ years, where $\tau_{BS}$ is the lifetime of the merger
product, a formation scenario is likely only if the derived encounter
time-scale is shorter than $\tau_{BS}$.  
Conversely, if the derived encounter time-scale is longer than 
$\tau_{BS}$, then that dynamical formation scenario
is unlikely to have occurred.  In
general, the shorter the derived encounter time-scale, the
more likely it is that one or more such encounters actually took place
within the lifetime of the merger product(s).  Finally, if the
derived encounter time-scale is longer than $\tau_{BS}$ for every
possible dynamical formation scenario, then a dynamical origin is
altogether unlikely for an observed multiple star system containing
one or more BSs.  Alternatively, the encounter time-scales must have
been shorter in the recent past. 
%

\section{Results} \label{results}

In this section, we apply our technique to two particular
cases of observed binaries and triples containing merger products.
The first is an observed triple system in the old open cluster M67
that is thought to contain two BSs \citep{vandenberg01,
  sandquist03}.  The second is the
period distribution of the BS binary
population of the OC NGC 188, which bears a remarkable
resemblance to M67 \citep{mathieu09}.  

\subsection{The Case of S1082} \label{s1082}

S1082 is believed to be a triple system in the old OC M67
\citep{vandenberg01, sandquist03}.  The observations suggest that a
distant triple companion orbits a close binary containing a BS and
another peculiar star.  The companion to the BS has a photometric
appearance that puts it close to the MSTO in the CMD.  And yet,
curiously, its derived mass is significantly greater than that of the
turn-off.  The outer companion is a BS in its own right, so that S1082
is thought to be composed of two BSs.  Although both the
inner and outer components of this suspected triple have systemic
velocites that suggest they are both cluster members, it is important
to note that there is no direct evidence proving a dynamical link
between the two \citep{sandquist03}.  Assuming for the time being that
a dynamical link does exist, we can apply the procedure described in
Section~\ref{method} to the case of S1082.

An orbital solution for S1082 has been provided by 
\cite{sandquist03}.  
By adding up the component masses, we find that the total mass of
S1082 is sufficiently large that its formation must have involved at
least 6 stars.  Based on this and a comparison of the different 
encounter time-scales, we find that the most probable formation
scenario for S1082 was a single 3+3 encounter.  However, we expect
very few, if any, 3+3 encounters to have occurred in the lifetime of a
typical merger product.  Therefore, our results are indicative of a low
probability of a system such as S1082 having formed dynamically in
M67.  On the other hand, if M67 had a
higher central density in the recent past, this would increase
the total encounter frequency.  In turn, this could significantly
increase the probability that S1082 has a dynamical origin.  We will
discuss this further in Section~\ref{discussion}.

\subsection{The Period-Eccentricity Distribution of the Blue Straggler
  Binary Population in NGC 188}

\cite{mathieu09} found 21 blue stragglers in the old open cluster NGC
188.  Of these, 16 are known to have a binary companion.  Orbital
solutions have been found for 15 of these known BS binaries.  From
this, \cite{mathieu09} showed that the BS binary population in NGC
188 has a curious
period-eccentricity distribution, with all but 3 having periods of
$\sim 1000$ days (we will call these long-period BS binaries).  Of
these three, two have periods of $< 10$ 
days (binaries 5078 and 7782; we will call these short-period BS
binaries).  The normal MS binary population, on the other
hand, shows
no sign of a period gap for $10 < P < 1000$ days
\citep{mathieu09}.  We can apply the procedure outlined in
Section~\ref{method} to better understand how we expect mergers
formed during dynamical encounters to contribute to the BS binary
population in NGC 188.  Although the method described in
Section~\ref{method} treats one system at a time, we will apply our
technique to the BS binary population of NGC 188 as a whole.

Before applying our technique, we must satisfy ourselves
that a dynamical origin is possible for a large fraction of the
observed BS population.  Analytic estimates for the times between the
different types of encounters suggest that a sufficient number of
suitable dynamical interactions 
should have occurred in the last $\tau_{BS}$ years for the formation
of every BS binary in NGC 188 to have been directly mediated by the
cluster dynamics.  

The first step of our technique is to find the
most commonly occurring type(s) of encounter(s).  Using estimates for
the binary and triple fractions taken from \cite{latham05},
Figure~\ref{fig:fb-ft} suggests that 1+2 encounters currently dominate
in NGC 188.  However, both 2+2 and 1+3 
encounters also occur with very comparable rates.  It follows that 1+3
encounters are the most important for BS formation in NGC 188 since
these are the most likely to result in stellar mergers.  This is
because most 2+2 and (especially) 1+2 encounters will not involve a
very hard binary as a result of their small cross sections for
collision.  On the other hand, in order for triples to remain dynamically
stable, the ratio between their outer and inner orbital semi-major
axes must be large \citep{mardling01}.  Most stable triples will
therefore contain a very hard inner binary and a triple companion on a
wide outer orbit.  The times between 1+2, 2+2 and 1+3
encounters are all comparable, suggesting that most encounters
involving hard binaries are 1+3 encounters.  These are also the most
important for BS formation since encounters involving very hard
binaries are the most likely to result in mergers \citep{fregeau04}.

The next step is to apply our energy conservation prescription
to the observed BS population.  In order to do this, we will consider
the short- and long-period BS binaries separately.  
Energy conservation suggests that the short-period BS binaries could
have formed from a 
direct stellar collision that occurred within a dynamical
encounter of a hard binary and another single or (hard) binary star
(we call this Mechanism I).  If at least one of the objects going into
the encounter was a triple, then four or more stars were involved in
the interaction.  Therefore, if binaries 5078
and 7782 were formed from this mechanism, they could
possess triple companions with sufficiently long periods that they
would have thus far evaded detection.  This is consistent with the
requirements for both conservation of energy and angular momentum.
Interestingly, the presence of an outer triple companion could also
contribute to hardening these BS
binaries via Kozai cycles operating in conjunction with tidal friction
\citep{fabrycky07}.

Now let us consider the long-period BS binaries.  From energy
conservation, we expect encounters involving
3 or more stars and only one very hard binary to typically produce
long-period BS binaries if the hard binary is driven to merge during
the encounter.  This is because the hard binary merges so that its
significant (negative) orbital energy can only be re-distributed to
the other stars by giving them positive energy.  Since the only other
orbits involved in the interaction are wide, the left-over BS binary
should also have a wide orbit.  Alternatively, BSs formed during
interactions involving more than one very hard
binary should be left in a wide binary provided enough energy is
extracted from the orbit of the binary that merges.  In this case, a
significant
fraction of this energy must be imparted to the other stars in order
to counter-balance the significant orbital energy of the other very
hard binary.  If not, the other short-period
binaries are required to either merge or be ejected from the
system.  Otherwise
energy conservation suggests that the left-over BS
binary should be very hard.  However, it is important to recall that
we
are neglecting other non-dynamical mechanisms for energy extraction.
We will return to this last point in Section~\ref{discussion}.

In order to help us obtain more quantitative estimates for the
long-period BS binaries, consider two additional
mechanisms for mergers during dynamical interactions that involve both
short- and long-period orbits.  First, a
merger can occur if a sufficient amount of orbital energy is extracted
from a hard binary by other interacting stars (called Mechanism IIa).
Alternatively, a merger can occur if the encounter progresses in such
a way that the eccentricity of a hard binary becomes sufficiently
large that the stellar radii overlap, causing the stars to
collide and merge (called Mechanism IIb).  In the case of Mechanism
IIb, most of the orbital energy of the close binary must end
up in the form of internal and gravitational binding energy in the
merger remnant after the majority of its orbital angular momentum has
been
redistributed to the other stars (and tides have extracted orbital
energy).  In the case of Mechanism IIa, however, most
of the orbital energy of the close binary must be imparted to
one or more of the other interacting stars in the form of bulk kinetic
motion.  Consequently, one or more stars are likely to obtain a
positive total energy and escape the system.  This need not
necessarily be the case for 2+3 and 3+3 encounters provided the second
hardest binary orbit involved in the interaction has a sufficiently
negative energy.

\begin{figure}
  \includegraphics[height=.3\textheight]{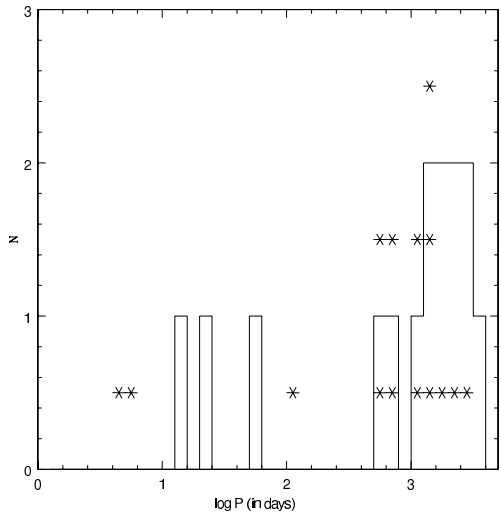}
\caption{Histogram of the period distribution (in
  days) expected for BS binaries formed during 15 1+3 encounters.
  The parameter space assumed for the encounters is described in the
  text.  The stars show the observed BS binary period
  distribution in NGC 188 taken from \cite{mathieu09}, where each
  star represents a single BS binary.}
\label{fig:per-num}
\end{figure}

Interestingly, the two general qualitative merger scenarios described
above
(Mechanisms I and II) naturally create a bi-modal period distribution
similar to the period gap observed for the BS binaries if we assume
that 1+3 encounters produced these objects.  To illustrate this,
Figure~\ref{fig:per-num} shows a histogram of periods for 15 BS
binaries formed during 1+3 encounters via these two generic merger
scenarios.  In order to obtain this plot, we have
used the observed period distribution for the regular
MS-MS binary population in NGC 188 from \cite{geller09} to obtain
estimates for the orbital energies of any binaries and/or triples
going into encounters.  Specifically, in order to obtain periods for
the outer orbits of triples undergoing encounters, we randomly sampled
the regular period distribution, including only those
binaries with periods satisfying $400$ days $< P < 4000$ days.  We
have argued that the initial outer orbits of triples going into 1+3
encounters provide a rough lower limit for the periods of BS binaries
formed via Mechanism II.  Therefore, any BS binaries formed in this
way could only have been identified as binaries by radial velocity
surveys if the triple going into the encounter had a period $< 4000$
days (since this corresponds to the current cut-off for detection).
All triples are taken to have a
ratio of 30 between their inner and outer orbital semi-major axes.
This ratio has been chosen to be arbitrarily large enough that the
triples should be dynamically stable.

We will adopt a ratio based on the observations of
\cite{mathieu09} for the fraction of outcomes that result in each of
these two possible merger scenarios.  In particular, if we assume that
the three BS
binaries with $P < 150$ days were formed via Mechanism I whereas the
other 12 were formed via Mechanism II (either IIa or IIb since energy
conservation predicts similar periods for the left-over BS binaries),
this would suggest that Mechanism II is $\sim 4$ times more likely to
occur than Mechanism I during any given 1+3 encounter.  The important
point to take away is that the observed BS
binary period-eccentricity distribution offers a potential constraint
on the fraction of encounters that result in different merger
scenarios.

\section{Summary \& Discussion} \label{discussion}

In these proceedings, we have presented a generalized prescription
for energy conservation during stellar encounters.  Our method can be
used to identify the most probable
dynamical formation scenario for an observed binary or triple
system containing one or more merger products.  We have shown
that, using the observed
orbital parameters of the system, the allowed initial orbital
semi-major axes of any binary or triple
systems involved in its formation can be constrained.  The
initial semi-major axes of the orbits in turn provide an
estimate for the collisional cross section and therefore the
time-scale
for the encounter to occur in its host cluster.  In order to apply our
technique, repeated spectroscopic measurements of the binary or triple
system containing the merger product(s) are needed in order to obtain
its orbital solution and systemic velocity.
However, the analytic time-scales can still
be applied if only the fraction of binary and triple stars are known,
which can be determined either spectroscopically
\citep[e.g.][]{mathieu90, latham05} or photometrically
\citep[e.g.][]{fan96}.

As we have shown, consideration of the requirement for energy
conservation is ideal for identifying trends during
stellar encounters, whereas
numerical scattering experiments can require hundreds or even
thousands of
simulation runs before patterns will emerge.  Some of these trends
include:

\begin{itemize}

\item At least one short-period binary is usually required in a
  dynamical interaction to produce another binary having a similarly
  short-period (provided no stars are ejected with escape velocities
  $> 100$ km/s).  This is because the orbital energy of a
  short-period binary is sufficiently negative that it tends to
  considerably outweigh the other forms of energy for most of the
  encounters that 
  typically occur in the cores of globular and especially open
  clusters.  This has been confirmed by \cite{hurley05}.

\item Previous studies have found that in order for triples to remain
  stable for many dynamical times, the ratio of their inner to outer
  orbital periods must be relatively large (roughly a factor of ten or
  more) \citep[e.g.][]{mardling01}.  Based on our results, this has
  two important corollaries for stellar mergers in dense cluster
  environments hosting a significant population of triples:

\begin{enumerate}

\item  The longest-lived triples will contain very hard inner binaries
with a significant orbital energy.  This is important since stellar radii are
in general more likely to overlap and hence mergers to occur
during resonant interactions involving very hard binaries
\citep[e.g.][]{fregeau04, hurley05}.  We therefore expect stellar
mergers
to be common during encounters involving stable triple systems.

\item  The longest-lived triples will contain wide outer orbits,
  creating
a large cross section for collision.  This suggests that the
time-scale required for a stable triple system to encounter another
object is typically short relative to the cluster age in dense
environments.  A significant fraction of encounters
involving very hard binaries, and hence resulting in stellar mergers,
will therefore involve triples in old open clusters such as M67
and NGC 188.

\end{enumerate}

\end{itemize}

\cite{vandenberg01} and \cite{sandquist03} suggest that
back-to-back binary-binary encounters, or even a single 3+3 encounter,
could have formed S1082.  We have
improved upon these previous studies by comparing the time-scales
required
for possible dynamical formation scenarios to occur.
Since we have argued in Section~\ref{results} that the
formation of S1082 must have involved at least 6 stars, it follows
that
only a 3+3 interaction could have reproduced the observed
configuration
via a single encounter.  However, the derived 3+3 encounter time-scale
is sufficiently long that we expect very few, if any, 3+3 encounters
occurred within the lifetime of a typical merger product.  Moreover,
the times for multiple encounters to occur are
longer than the cluster age.  Although we cannot rule out a dynamical
origin for S1082, our results suggest that it is unlikely (provided
the
derived encounter time-scales were not significantly higher in the
recent past, which we will return to below).  From this, it follows
that a
dynamical link between the close binary and third star is unlikely to
exist.

On the other hand, we have so far ignored the cluster evolution,
and assumed that the currently observed cluster parameters have not
changed in the last few Gyrs.  N-body models suggest that the central
density in M67 could have been significantly higher in the recent
past.  Specifically, Figure 5 of \cite{hurley05} indicates that the
presently observed central density in M67 could have been higher
within the past Gyr by a factor of $> 2$.  If this was indeed
the case, our previous estimates for each of the different encounter
frequencies should increase by a factor of $\sim 4$, so that a
significant number of dynamical encounters involving single, binary
and triple stars should have occurred in M67 within the last
$\tau_{BS}$ years.  It follows that a dynamical
origin for S1082 is not unlikely if the central density in M67 was
recently larger than its presently observed value by a factor of
$> 2$.  This also increases the
probability that a scenario involving multiple encounters created
S1082, although we have shown that such a scenario is still likely to
have involved one or more triples.

Based on the preceding arguments, S1082 offers an excellent example of
how observations of
individual multiple star systems containing BSs can be used to
directly constrain the dynamical history of their host cluster.
If a definitive dynamical link between components A and B is
established, this would suggest that the central density in M67
was higher in the last 1-2 Gyrs.  This is also required in order
for the cluster dynamics to have played a role in the formation of a
significant fraction of the observed BS population in M67.  Based on
the current density, the encounter time-scales are sufficiently long
that too few encounters should have occurred in the last $\tau_{BS}$
years for mergers during dynamical interactions to be a significant
contributor to BS formation.

We have obtained quantitative constraints
for two generic channels for mergers during encounters involving
triples -- one in which a direct stellar collision occurs within a
dynamical interaction of the hard inner binary of a triple and another
single or (hard) binary star (Mechanism I) and
one in which the hard inner binary of a triple is driven to coalesce
by imparting energy and/or angular momentum to other stars involved in
the
interaction (Mechanism II).  Our results
suggest that these two general merger mechanisms could contribute to a
bi-modal period distribution for BS binaries similar to that
observed in NGC 188.  These dual mechanisms predict a gap in
period, with those BS binaries formed via Mechanism I having periods
of a
few to $\sim 100$ days and those formed via Mechanism II having
periods closer to $\sim 1000$ days.  Some 2+2, 2+3 and even 1+3
encounters could involve orbits with periods in this range, and
energy conservation confirms that the final period of
a BS binary formed via Mechanism II will typically be determined by
that of the second hardest binary orbit.  Therefore, we might still
expect some BS binaries to have periods that fall in the gap (100 days
$<$ P $<$ 1000 days).  Our results do indeed predict one
such BS binary, as shown in Figure~\ref{fig:per-num}.

Our results highlight the need for simulations of
1+3, 2+3 and 3+3 encounters to be performed in order to better
understand their expected contributions to BS populations in open and
globular clusters.  Once a preferred encounter scenario has been
identified for an observed binary or triple containing one or more
BSs, numerical scattering experiments can be used to
further constrain the conditions under which that scenario will occur
(or to show
that it cannot occur).  We have demonstrated that a combination of
observational and analytic constraints can be used to isolate the
parameter space relevant to the dynamical formation of an observed
multiple star system (or population of star systems) containing
one or more merger products.  This will drastically narrow the
relevant parameter space for numerical scattering experiments.

We have improved upon the results of \cite{perets09} and
\cite{mathieu09} since we have shown that dynamical encounters
involving triples could not only be contributing to
the long-period BS binaries in NGC 188, but they could also be an
important formation
mechanism for short-period BS binaries and triples containing BSs.  We
have not ruled out mass transfer or Kozai-induced
mergers in triples (primordial or otherwise) \citep{mathieu09,
  perets09}, or even various
combinations of different mechanisms, as contributing formation
channels
to the BS binary population in NGC 188.  For instance, a 1+3 exchange
interaction could stimulate a merger indirectly if the resulting angle
of inclination between the inner and
outer orbits of the triple exceeds $\sim 39^{\circ}$, ultimately
allowing
the triple to evolve via the Kozai mechanism so that the eccentricity
of the inner binary increases while its period remains roughly
constant \citep{eggleton06}.  For these reasons,
a better understanding of triple evolution, as well as binary
evolution in binaries containing merger products, is needed.

To conclude, we have presented an analytic technique to constrain the dynamical
origins of multiple star systems containing one or more BSs.  Our
results suggest that, in old open clusters, most dynamical
interactions resulting in mergers involve triple stars.  If most
triples are formed dynamically, this could suggest that
many stellar mergers are the culmination of a hierarchical build-up of
dynamical interactions.
Consequently, this mechanism for BS formation should
be properly included in future N-body simulations of cluster
evolution.  A better
understanding of the interplay between the cluster dynamics and the
internal evolution of triple systems is needed in order to better
understand the expected period distribution of BS binaries formed from
triples.  Simulations will therefore need to
track both the formation
and destruction of triples as well as their internal evolution via
Kozai cycles, stellar and binary evolution, etc.  On the observational
front, our results highlight the need for a more detailed knowledge of
binary and especially triple populations in clusters.

\begin{theacknowledgments}
We would like to thank Bob Mathieu for many helpful comments and
suggestions.  We would also like to thank Aaron Geller, Evert
Glebbeek, Hagai Perets, David Latham, Daniel Fabrycky and Maureen van
den Berg for useful discussions.  This research has been supported by
NSERC and OGS as well as the National Science Foundation under Grant
No. PHY05-51164 to the Kavli Institute for Theoretical Physics.
\end{theacknowledgments}

\bibliographystyle{aipproc}   
\bibliography{sample}


\end{document}